\documentclass[12pt]{article}

\usepackage[notref,notcite]{showkeys}
\usepackage{epsf,amsfonts,hyperref,macros}
\usepackage{cite}

\newcommand{\sgn} {{\rm sgn\,}}

\newcommand{\Dmu} {\,\overline{\rm d} }

\begin{document}
%=============================Page de titre===============
%\date{??}
%\author{Eynard}
%\title{<TITRE>}
%\topmargin .5cm \textheight 21.5cm \textwidth 15.8cm 
%\oddsidemargin 0.54cm
%\evensidemargin 0.54cm 
\sloppy

%\maketitle

\pagestyle{empty}
\hfill SPT-00/170, CRM 2698
\addtolength{\baselineskip}{0.20\baselineskip}
\begin{center}
\vspace{26pt}
{\large \bf {Asymptotics of skew orthogonal polynomials}}
\newline
\vspace{26pt}

{\sl B.\ Eynard}\hspace*{0.05cm}\footnote{ E-mail: eynard@spht.saclay.cea.fr }\\
\vspace{6pt}
Service de Physique Th\'{e}orique de Saclay,\\
F-91191 Gif-sur-Yvette Cedex, France.\\
\end{center}
\vspace{20pt}
\begin{center}
{\bf Abstract}
\end{center}
%-----------------------------ABSTRACT--------------------------------------
%
Exact integral expressions of the skew orthogonal polynomials involved in 
Orthogonal ($\beta=1$) and Symplectic ($\beta=4$) random matrix ensembles are obtained: the (even rank) skew orthogonal polynomials are average characteristic polynomials of random matrices.
From there, asymptotics of the skew orthogonal polynomials are derived.

\newpage
\pagestyle{plain}
\setcounter{page}{1}

%*********************************************************************
%==================== ARTICLE ========================================
%*********************************************************************

%================================ Introduction =============================
\newsection{Introduction}
\label{intro}

Families of Orthogonal (or skew-orthogonal) Polynomials, have many applications 
to mathematics and physics \cite{SzegoBook,DeiftBook}.

Here, we will have in mind applications to Random Matrix Theory 
(RMT)\cite{MehtaBook,Guhr:1997ve,FGZ:1993fi,KunzBook,Verbaarschot:1997qm}, i.e. 
disordered 
solid state physics\cite{Guhr:1997ve}, QCD\cite{Verbaarschot:1997qm}, or 
statistical physics on a random fluctuating lattice\cite{FGZ:1993fi,BIPZ} 
(2D quantum gravity).
In all these fields of physics, one is interested in the spectrum of a matrix 
(Hamiltonian, Transmission matrix, S-matrix, Dirac operator,...), which can be 
considered as random for various reasons (disorder, random impurities, quantum 
fluctuations, chaos or non-integrability,...).
It was observed that the spectrum of a large random matrix shows universal 
properties\cite{Bohigas:1984, Pastur:1997} (2-point correlation function, in the 
short or long range regime;  
universal conductance fluctuations of mesoscopic conductors).
One possible way to understand and prove that universality is through the 
``orthogonal 
polynomials'' method, which we shall recall below.
In order to extract some useful numerical results, it is important to have some 
asymptotics of the orthogonal polynomials in some special limit.

The type of orthogonal polynomials involved, depends on the symmetry of the 
matrix ensemble \cite{MehtaBook}.
The case of a physical system with broken time-reversibility (for instance a 
mesoscopic conductor in the presence of a magnetic field), represented by a 
$U(N)$ invariant matrix ensemble, was extensively studied, because it is the 
simplest\cite{DeiftBook}.

Here, we shall focus on the $O(N)$ and $Sp(2N)$ invariant matrix ensembles, 
which appear for physical systems with time-reversibility and/or half-integer 
spin with broken rotational symmetry.
These ensembles involve families of skew-orthogonal polynomials.

The aim of this article is to present a remarkable exact expression of the skew 
orthogonal 
polynomial as an integral, and deduce from it the required asymptotics.

\bigskip
Section 2 is a brief introduction to the orthogonal polynomial's method in RMT, 
in section 3 we give and prove the remarkable exact expressions for the 
skew-orthogonal polynomials, and in section 4, we consider their asymptotics.

\newsection{The Orthogonal Polynomials}

Consider the partition function of a random matrix $M$:

\beq\label{Zdef}
Z^{(\beta)}_N[V] = \int_{M\in E^{(\beta)}_N} \D{M} \,\, \ee{-{N_\beta} \tr V(M) 
}
\eeq
where $E^{(1)}_N$ is the set of all $N\times N$ real symmetric matrices, 
$E^{(2)}_N$ is the set of all $N\times N$ hermitian matrices, $E^{(4)}_N$ is the 
set of all $2N\times 2N$ self-adjoint real quaternionic matrices\footnote{$M\in 
E^{(4)}_N$ can be viewed either as a $2N\times 2N$ matrix with complex number 
entries or a $N\times N$ matrix with quaternion entries. It has $N$ eigenvalues, 
each twice degenerated.\cite{MehtaBook}}, and $\D{M}$ is the Haar measure on $ 
E^{(\beta)}_N$.
$V(x)$ is a polynomial potential, bounded from below, and $N_\beta= N,N, N/2$ 
respectively for $\beta=1,2,4$.

The angular degrees of freedom of $M$ can be integrated out, and \ref{Zdef} can 
be rewritten as an integral over the $N$ eigenvalues $(\l_1,\dots,\l_{N})$ of 
$M$ only\cite{Mehta:1981,MehtaBook}:

\beq\label{Zvpdef}
Z^{(\beta)}_N[V] = U^{(\beta)}_N \int \prod_{i=1}^{N} \Dmu{\l_i} \,\,\, 
|\Delta(\l)|^\beta 
\eeq
where
$U^{(\beta)}_N$ is the volume of the group $O(N)$, $U(N)$ or $Sp(2N)$ 
respectively for $\beta=1,2,4$.
$\Dmu{\l} = \D{\l}\, \ee{-N V(\l)} $ is the measure element,
and
\beq
\Delta(\l) = \prod_{i<j} (\l_i-\l_j)
\eeq
is the Vandermonde determinant, which can be rewritten as:
\beq\label{VDMdef}
\Delta(\l) = \det {\pmatrix{1 & \l_1 & \l_1^2 & \dots & \l_1^{N-1} \cr 1 & \l_2 
& \l_2^2 & \dots & \l_2^{N-1} \cr  \vdots &  &  &  & \vdots \cr 1 & \l_{N} & 
\l_{N}^2 & \dots & \l_{N}^{{N}-1} \cr  } } = \det (\l_i^j) = \det P_j(\l_i)
\eeq
where $P_j(\l) = \l^j+\dots$ is an arbitrary monic polynomial of degree $j$.
The last equality is obtained by linearly mixing columns of the determinant, and 
the first equality is the well known Vandermonde determinant, which can be found 
in any math textbook\cite{MehtaBook}.

The computation of integral \ref{Zvpdef} becomes easier with a special 
choice of the polynomials $P_j(\l)$, chosen orthogonal with respect to an 
appropriate scalar product \cite{Mehta:1981}:

\begin{itemize}
\item In the unitary case $\beta=2$, the scalar product under consideration is:
\beq\label{scalarU}
<f|g> = \int_{-\infty}^\infty \, \Dmu{x} \, f(x) g(x) 
\eeq
and the polynomials $P_n(x)$ are chosen orthogonal:
\beq
<P_n | P_m > = h_n \delta_{nm}
\eeq

\item  In the orthogonal case $\beta=1$, the scalar product under consideration 
is skew-symmetric:
\beq\label{scalarO}
<f|g> = - <g|f> = \int_{-\infty}^\infty \int_{-\infty}^\infty \, \Dmu{x}\Dmu{y} 
\,\,\, f(x)\,\,  \sgn{(x-y)}\,\, g(y)
\eeq
and the polynomials $P_n(x)$ are chosen skew-orthogonal:
\bea
<P_{2n} | P_{2m} > = <P_{2n+1} | P_{2m+1} >  = & 0 \\
<P_{2n+1} | P_{2m} > = &  h_n \delta_{nm} 
\eea

\item  In the symplectic case $\beta=4$, the scalar product under consideration 
is skew-symmetric too:
\beq\label{scalarSp}
<f|g> = - <g|f> = \int_{-\infty}^\infty  \, \Dmu{x} \, (f(x) g'(x) -f'(x)g(x) ) 
\eeq
and the polynomials $P_n(x)$ are chosen skew-orthogonal:
\bea
<P_{2n} | P_{2m} > = <P_{2n+1} | P_{2m+1} > = & 0 \\
<P_{2n+1} | P_{2m} >  = & h_n \delta_{nm}
\eea

\end{itemize}

In all three cases, the partition function \ref{Zvpdef} reduces 
mainly\footnote{The actual result may depend on the parity of $N$. Details can 
be found in \cite{}.} to $Z=\prod_{n=1}^{n_F} h_{n-1}$, where $n_F$ is 
called the "Fermi level" by analogy with a system of fermions:
\beq
n_F = {\beta\over 2} N_\beta \for \beta=(1,2,4)
\eeq
When $N$ is large, most of the physical quantities and relevant observables are related to properties of $h_n$ in the vicinity of the Fermi 
level: $n\to\infty$, $N\to\infty$ and $n-n_F \sim O(1)$.

\subsection{Determination of the orthogonal polynomials}

For a generic potential $V(x)$, those orthogonal polynomials exist, and 
can be constructed by recurrence.
Indeed, we start from $ P_0(x) = 1 $,
then the coefficients of $P_1$ are determined by the orthogonality conditions, 
and by recurrence, we determine $P_n$ and $h_n$ for all $n$.

Note that for the skew-orthogonal polynomials, there is an ambiguity: 
$P_{2n+1}$ is defined only up to an arbitrary linear combination with $P_{2n}$.
If one wants a unique definition, an extra condition should be added, for 
instance that the term of degree $2n$ in $P_{2n+1}$ vanishes.
Anyway, the values of $h_n$ don't depend on this ambiguity.

\medskip

The determination of the orthogonal polynomials by recurrence is unefficient if 
one wants to compute $P_n$ for $n$ large.
The aim of this article is to present a closed expression of $P_n$ for any $n$, 
and to derive from it some asymptotics in the large $n$ limit, and particularly 
near the Fermi level $n-n_F \sim O(1)$.

\section{An exact expression of the skew-orthogonal polynomials}

\begin{itemize}
\item In the unitary case $\beta=2$, it is known that

\bea\label{PnUint}
P^{(2)}_n(x) & = & {1\over Z^{(2)}_n} \int_{M\in E^{(2)}_n} \D{M} \, \det{(x-M)} 
\,\, \ee{-N\tr V(M)} \\
\nonumber & = & {U_n^{(2)}\over Z^{(2)}_n} \int \Dmu{\l_1}\dots \Dmu{\l_n} \, 
\prod_{i<j} 
|\l_i-\l_j|^2 \, \prod_i (x-\l_i) 
\eea
The $n^{\rm th}$ orthogonal polynomial is the average of the characteristic polynomial of a 
$n\times n$ hermitian matrix with respect to the weight $\ee{-N\Tr V(M)}$:
\beq\label{PnUexact}
P_n^{(2)}(x) = \left< \det{(x-M)} \right>_{n\times n}
\eeq

This has been known for more than a century \cite{Heine:1878} (in the context of RMT, see e.g. \cite{SzegoBook,DeiftBook,Eynard:1997jf}).
We are now going to generalize this expression to $\beta=1$ and $4$.

\item Orthogonal case $\beta=1$.
We will prove below that:

\bea\label{PnOint}
P^{(1)}_{2n}(x) & = & {1\over Z^{(1)}_{2n}}\int_{M\in E^{(1)}_{2n}} \D{M} \, 
\det{(x-M)} \,\, \ee{-N\tr V(M)} \\
\nonumber & = & {U_{2n}^{(1)}\over Z^{(1)}_{2n}} \int \Dmu{\l_1}\dots 
\Dmu{\l_{2n}} \, 
\prod_{i<j} |\l_i-\l_j| \, \prod_i (x-\l_i) \\
\nonumber   & = & \left< \det{(x-M)} \right>_{2n\times 2n} \\
\nonumber {\rm and} & & \\
P^{(1)}_{2n+1}(x) & = & {1\over Z^{(1)}_{2n}}\int_{M\in E^{(1)}_{2n}} \D{M} \, 
(x+\tr M + c_n ) \det{(x-M)} \,\, \ee{-N\tr V(M)} \\
\nonumber  & = & {U_{2n}^{(1)}\over Z^{(1)}_{2n}} \int \Dmu{\l_1}\dots 
\Dmu{\l_{2n}} \, 
\prod_{i<j} |\l_i-\l_j| \,\, (x+\sum_i \l_i + c_n ) \,\prod_i (x-\l_i)  \\
\nonumber   & = & \left< (x+\Tr M + c_n)\det{(x-M)} \right>_{2n\times 2n} 
\eea
the constants $c_n$ can be chosen arbitrarily, the choice $c_n=0$ is the one 
such that the term of degree $2n$ in $P_{2n+1}$ vanishes.

\item Symplectic case $\beta=4$

\bea\label{PnSpint}
P^{(4)}_{2n}(x) & = & {1\over Z^{(4)}_{n}}\int_{M\in E^{(4)}_n} \D{M} \, 
\det{(x-M)} \,\, \ee{-{N\over 2}\tr V(M)} \\
\nonumber & = & {U_n^{(4)}\over Z^{(4)}_{n}} \int \Dmu{\l_1}\dots \Dmu{\l_{n}} 
\, 
\prod_{i<j} |\l_i-\l_j|^4 \, \prod_i (x-\l_i)^2 \\
\nonumber   & = & \left< \det{(x-M)} \right>_{n\times n} \\
\nonumber {\rm and} & & \\
P^{(4)}_{2n+1}(x) & = & {1\over Z^{(4)}_{n}}\int_{M\in E^{(4)}_n} \D{M} \, 
(x+\tr M + c_n )\det{(x-M)} \,\, \ee{-{N\over 2}\tr V(M)} \\
\nonumber & = & {U_n^{(4)}\over Z^{(4)}_{n}} \int \Dmu{\l_1}\dots \Dmu{\l_{n}} 
\, 
\prod_{i<j} |\l_i-\l_j|^4 \,\, (x+2\sum_i \l_i + c_n ) \,\prod_i (x-\l_i)^2 \\
\nonumber   & = & \left< (x+\Tr M + c_n)\det{(x-M)} \right>_{n\times n} 
\eea

\end{itemize}

\subsection{Proof of \ref{PnOint} }

Note that it is sufficient to prove that 
\beq
<P_{2n} | x^m > = 0 \qquad {\rm and} \qquad  <P_{2n+1} | x^m > = 0 \qquad {\rm 
for \,\, all} \quad m\leq 2n-1
\eeq

Consider:
\bea
 <P_{2n} | x^m >  & \propto &  \int  \Dmu{x}\,\Dmu{y}\, \Dmu{\l_1}\dots 
\Dmu{\l_{2n}}\,\,  \\
\nonumber & &   \prod_{i<j} (\l_i-\l_j) \,\, \prod_i (x-\l_i) \,\, \prod_{i<j}\sgn{(\l_i-\l_j)}\,\, \sgn{(x-y)} \,\,\, y^m 
\eea
then write $x=\l_{2n+1}$:
\bea
<P_{2n} | x^m > & \,\propto  \int & \Dmu{y}\, \Dmu{\l_1}\dots 
\Dmu{\l_{2n+1}}\,\,   \,\, y^m \, \prod_{1\leq i<j\leq 2n+1} (\l_i-\l_j)  \\
\nonumber & & \prod_{1\leq i<j\leq 2n+1} \sgn{(\l_i-\l_j)}\,\,  \prod_{i=1}^{2n} 
\sgn{(\l_i-\l_{2n+1})}\,\,\sgn{(\l_{2n+1}-y)} \, 
\eea
symmetrize with respect to the first $2n+1$ variables:
\bea
<P_{2n} | x^m > & \,\propto & \sum_{k=1}^{2n+1} \int  \Dmu{y}\, \Dmu{\l_1}\dots 
\Dmu{\l_{2n+1}}\,\,\,  y^m  \,\,\prod_{1\leq i<j\leq 2n+1} (\l_i-\l_j)  \\
\nonumber  & & \prod_{1\leq i<j\leq 2n+1} \sgn{(\l_i-\l_j)}\,\,  \prod_{1\leq i 
\neq k \leq 2n+1} \sgn{(\l_i-\l_{k})}\,\,\sgn{(\l_{k}-y)}  
\eea
Note the following identity:
\beq
\prod_{i=1}^{2n+1} \sgn{(y-\l_i)} =  \sum_{k=1}^{2n+1} \sgn(y-\l_k)  \prod_{i=1, 
i\neq k}^{2n+1} \sgn(\l_{k}-\l_i)
\eeq
which gives (and note $y=\l_{2n+2}$):
\bea
\nonumber <P_{2n} | x^m >  \,\propto & \int  \Dmu{\l_1}\dots 
\Dmu{\l_{2n+2}}\,\,\, \left[ \l_{2n+2}^m \,\, \prod_{1\leq i<j\leq 2n+1} 
(\l_i-\l_j)  \right]  \\
 & 
\left[ \prod_{1\leq i<j\leq 2n+2} \sgn{(\l_i-\l_j)} \right]
\eea
The second bracket is completely antisymmetric in the $2n+2$ variables, so that 
we have to antisymmetrize the first bracket as well.
The result is zero when $m\leq 2n$, because any non-zero antisymmetric 
polynomial of $2n+2$ variables must have degree at least $2n+1$, while the first bracket is a polynomial of degree at most $2n$ in any of its variables.

By the same argument, one would find that
\bea
\nonumber <P_{2n+1} | x^m >  \,\propto & \int  \Dmu{\l_1}\dots 
\Dmu{\l_{2n+2}}\,\,\, \left[ \l_{2n+2}^m \,\, (c_n + \sum_{i=1}^{2n+1} \l_i ) 
\,\, \prod_{1\leq i<j\leq 2n+1} (\l_i-\l_j)  \right]  \\
 & 
\left[ \prod_{1\leq i<j\leq 2n+2} \sgn{(\l_i-\l_j)} \right]
\eea
which, by antisymmetrization of the first bracket, vanishes when $m\leq 2n-1$.

\subsection{Proof of \ref{PnSpint} }

Again, it is sufficient to prove that 
\beq
<P_{2n} | x^m > = 0 \qquad {\rm and} \qquad  <P_{2n+1} | x^m > = 0 \qquad {\rm 
for \,\, all} \quad m\leq 2n-1
\eeq

\bea\label{Pspxm}
<P_{2n} | x^m >  & \propto & \int \Dmu{x}\,\Dmu{\l_1}\dots \Dmu{\l_{n}} \\
\nonumber & & \,\,\,
\prod_{1\leq i<j\leq n} (\l_i-\l_j)^4 \prod_{i=1}^n (x-\l_i)^2  \,\,\, \left( m 
x^{m-1} -  x^m \sum_{i=1}^{n} {2\over x-\l_i} \right)
\eea
Introduce $n$ extra variables $(\mu_1, \dots , \mu_n)$, and consider the 
$2n\times 2n$ Vandermonde determinant of the $2n$ variables $(\l_i, \mu_i)$, 
divide it by $\prod_i (\l_i-\mu_i)$ and take the limit $\mu_i \to \l_i$.
You get:
\beq
\prod_{1\leq i<j\leq n} (\l_i-\l_j)^4 = \mathop{\rm lim}_{\mu_i \to \l_i} {  
\Delta_{2n}(\l_i,\mu_i)  \over \prod_{i=1}^{n} (\l_i-\mu_i)}
= \det 
\pmatrix{
1 & \l_1 & \l_1^2 & \dots & \l_1^{2n-1} \cr
\vdots  &  &  &  & \vdots  \cr
1 & \l_n & \l_n^2 & \dots & \l_n^{2n-1} \cr 
0 & 1 & 2\l_1 & \dots & (2n-1)\l_1^{2n-2} \cr 
\vdots  &  &  &  & \vdots  \cr
0 & 1 & 2\l_n & \dots & (2n-1)\l_n^{2n-2} \cr 
}
\eeq

With the same trick, we have:

\bea
\prod_{1\leq i<j\leq n} (\l_i-\l_j)^4 \prod_{i=1}^n (x-\l_i)^2
 & = & \mathop{\rm lim}_{\mu_i \to \l_i}  {  \Delta_(x,\l_i,\mu_i)  \over \prod_{i=1}^n (\l_i-\mu_i)} \,\, = \\
\nonumber & = & \det 
\pmatrix{
1 & x & x^2 & \dots & x^{2n} \cr
1 & \l_1 & \l_1^2 & \dots & \l_1^{2n} \cr
\vdots  &  &  &  & \vdots  \cr
1 & \l_n & \l_n^2 & \dots & \l_n^{2n} \cr 
0 & 1 & 2\l_1 & \dots & 2n\l_1^{2n-1} \cr 
\vdots  &  &  &  & \vdots  \cr
0 & 1 & 2\l_n & \dots & 2n\l_n^{2n-1} \cr 
}
\eea

and
\beq
{\d\over \d x} \prod_{1\leq i<j\leq n} (\l_i-\l_j)^4 \prod_{i=1}^n (x-\l_i)^2 
 = \det 
\pmatrix{
1 & \l_1 & \l_1^2 & \dots & \l_1^{2n} \cr
\vdots  &  &  &  & \vdots  \cr
1 & \l_n & \l_n^2 & \dots & \l_n^{2n} \cr 
0 & 1 & 2x & \dots & (2n+1) x^{2n-1} \cr 
0 & 1 & 2\l_1 & \dots & 2n\l_1^{2n-1} \cr 
\vdots  &  &  &  & \vdots  \cr
0 & 1 & 2\l_n & \dots & 2n\l_n^{2n-1} \cr 
}
\eeq

Therefore, the integrand in \ref{Pspxm}  is a $(2n+2)\times (2n+2)$ determinant:
\beq
\det 
\pmatrix{
1 & x & x^2 & \dots & x^{2n} & x^m \cr
1 & \l_1 & \l_1^2 & \dots & \l_1^{2n} & 0 \cr
\vdots  &  &  &  & \vdots & \vdots  \cr
1 & \l_n & \l_n^2 & \dots & \l_n^{2n} & 0 \cr 
0 & 1 & 2x & \dots & (2n+1) x^{2n-1} & m x^{m-1} \cr 
0 & 1 & 2\l_1 & \dots & 2n\l_1^{2n-1} & 0 \cr 
\vdots  &  &  &  & \vdots & \vdots  \cr
0 & 1 & 2\l_n & \dots & 2n\l_n^{2n-1} & 0 \cr 
}
\eeq

we note $x=\l_{n+1}$, and by antisymmetrization, it becomes:
\beq
\det 
\pmatrix{
1 & \l_1 & \l_1^2 & \dots & \l_1^{2n} & \l_1^m \cr
\vdots  &  &  &  & \vdots & \vdots  \cr
1 & \l_{n+1} & \l_{n+1}^2 & \dots & \l_{n+1}^{2n} & \l_{n+1}^m \cr 
0 & 1 & 2\l_1 & \dots & 2n\l_1^{2n-1} & m\l_1^{m-1} \cr 
\vdots  &  &  &  & \vdots & \vdots  \cr
0 & 1 & 2\l_{n+1} & \dots & 2n\l_{n+1}^{2n-1} & m\l_{n+1}^{m-1} \cr 
}
\eeq
which obviously vanishes when $m\leq 2n$.

By the same argument, one would find that
$<P_{2n+1} | x^m > $ reduces to the same kind of integral, but with $m$ replaced 
by $m+1$, and vanishes when $m\leq 2n-1$.

\medskip

We have thus proven that the skew-orthogonal polynomials are indeed given by 
\ref{PnOint} and \ref{PnSpint}.

\section{Large $N$ asymptotics}

Most of the large $N$ universal statistical properties of a random $N\times N$ 
matrix $M$ belonging to one of the three ensembles $E^{(\beta)}_N$, can be 
expressed in terms of a few  $h_n$, with $n$ close to the Fermi level \beq
n_F={\beta\over 2} N_\beta
\eeq

More precisely, for $\beta=2$, we need asymptotics of $P_n$ in the limit
\beq\label{needasympU}
N\to\infty \virg n\to\infty \virg n-N \sim O(1) 
\eeq
for $\beta=1$, we need asymptotics of $P_{2n}$ and $P_{2n+1}$ in the limit
\beq\label{needasympO} 
N\to\infty \virg n\to\infty \virg 2n- N \sim O(1) 
\eeq
and for $\beta=4$, we need asymptotics of $P_{2n}$ and $P_{2n+1}$ in the limit
\beq\label{needasympSp} 
N\to\infty \virg n\to\infty \virg n- N =n-2 N_4 \sim O(1) 
\eeq

\subsection{The resolvant}

We introduce the function $W(z)$ usually called the resolvant or Green function:
\beq
W(z) := W^{(\beta)}_{m} [{\cal V}](z) :=  {1\over m} \left< \sum_{k=1}^{m} 
{1\over z-\l_k} \right> \,\, \propto \,\, {1\over m} \left< \tr {1\over z-M} 
\right>
\eeq
where $M\in E_m^{(\beta)}$ and the mean value is taken with respect to the 
weight:
\beq
\ee{-{m_\beta} \Tr {\cal V}(M)}
\eeq 
When there is no ambiguity, we will drop the $\beta$, $m$ or ${\cal V}$ indices, 
and write the resolvant as $W(z)$.
Note that we have chosen a normalization such that:
\beq\label{largezresolvant}
W(z) \mathop{\sim}_{z\to \infty} \, {1\over z}
\eeq

The reason to introduce the resolvant is that the logarithmic derivative of 
$P_n(x)$ is proportional to the resolvant $W_m(z)$ (from 
\ref{PnUint},\ref{PnOint},\ref{PnSpint}, at least when $n$ is even) for some 
appropriate value of $m$, and with a potential of the form:
\beq 
{\cal V}(z) =  {1\over T}V(z)-{r}\ln(x-z) .
\eeq

More precisely, we have:

\begin{itemize}
\item In the Unitary case $\beta=2$:
\beq
{  {P^{(2)}_{n}}'(x) \over  P^{(2)}_{n}(x) } = n
\left. W_n(z) \right|_{z=x} 
\with
{\cal V}(z) = {N\over n} V(z)-{1\over n}\ln{(x-z)}
\eeq
i.e. $m=n$, $r={1\over n}$ and $T={n\over N}$ ($\to 1$ when $n\to n_F$).

\item In the Orthogonal case $\beta=1$:

\beq
{{P^{(1)}_{2n}}'(x) \over P^{(1)}_{2n}(x)}  = 2n
\left. W_{2n}(z) \right|_{z=x}
\with
{\cal V}(x) = {N\over m} V-{1\over m}\ln(x-z)
\eeq
i.e. $m=2n$, $r={1\over2n}$ and $T={2n\over N}$ ($\to 1$ when $n\to n_F$).

\item In the Symplectic case $\beta=4$:

\beq
{{P^{(4)}_{2n}}'(x) \over P^{(4)}_{2n}(x)}  = 2n
\left. W_{n}(z) \right|_{z=x}
\with
{\cal V}(x) = {N\over n} V-{2\over n}\ln(x-z)
\eeq
i.e. $m=n$, $r={2\over n}$ and $T={n\over N}$ ($\to 1$ when $n\to n_F$).

\end{itemize}

In all three cases: $T={n\over n_F}$ and $r={\beta \over 2n}$.

\subsection{Asymptotics for the Resolvant}

In a potential ${\cal V}$, the resolvant $W(z)=W_m(z)$ satisfies the equations 
of motion (resulting from invariance of an integral like eq.\ref{Zdef} under a change of variable 
$M\to  f(M)$):
\beq\label{eqmotion}
 W(z)^2 - {\eta \over 2n} W'(z) = {2\over \beta} {\cal V}'(z) W(z) - Q(z) 
+O(1/ n^2)
\eeq
where $\eta=(1,0,-1)$ respectively for $\beta=(1,2,4)$ and $Q(z)$ is a 
polynomial\footnote{when ${\cal V}'$ has poles, $Q$ may have poles too. $Q(z)$ 
is a rational function, whose poles must be chosen in order to cancel the poles of $W(z)$ in eq.\ref{eqmotion}.} of degree $\deg V -2$, 
which is not determined by the equations of motions, it has to be determined by 
analytical considerations, for instance the one-cut asumption.

Here, we will consider a potential ${\cal V}$ of the form:
\beq
{\cal V}(z) = {1\over T} V(z) - r \ln{(x-z)}
\eeq
and we will be interested in the limit where $T-1$ and $r$ are small of order 
$1/n$.

The method is to find first the solution $W(z)$ at $T=1$ and $r=0$.
We write it:
\beq
W(z) = W_0(z) + {\eta\over 2n} W_1(z) + O(1/n^2)
\eeq
and then, add the variations:
\beq
(T-1) {\d\over \d T} W_0 + r {\d\over \d r} W_0
\eeq
(at order $1/n$, we don't need to consider the variations of $W_1$ with respect 
to $T$ and $r$), the derivatives are taken at $T=1$ and $r=0$.

\subsection{Contribution of $W_0$}

The function $W_0(z)$, (as well as its derivatives with respect to $T$ and $r$) 
has been extensively studied in RMT.
Note that $W_0$ is nearly the same for $\beta=1$, $2$ or $4$.
Let us recall here some of the main features of $W_0$ in order to fix the 
notations.

\bigskip

At $n\to\infty$ (and $T=1$ and $r=0$), \ref{eqmotion} reduces to a quadratic 
equation for $W_0(z)$.
The one-cut-solution is:
\beq\label{onecutsol}
 W_0(z)  = {1\over \beta}\left( V'(z) - M(z)\sqrt{(z-a)(z-b)} \right) = {1\over 
\beta} V'(z) - i\pi \rho(z)
\eeq
Where $M(z)$ is a polynomial of degree $d-1$ ($d=\deg V'$), which is completely 
determined by the large $z$ limit condition \ref{largezresolvant}:
\beq
 M(z) = {\rm Pol}\,\, {V'(z)\over \sqrt{(z-a)(z-b)}}
\eeq
The end-points $a$ and $b$ too, are determined by \ref{largezresolvant} which 
implies:
\beq \oint {V'(z)\over \sqrt{(z-a)(z-b)}} \D{z} = 0 \virg \oint {zV'(z)\over 
\sqrt{(z-a)(z-b)}} \D{z} = 2i\pi \beta
\eeq
where the contour encircles the cut $[a,b]$ in the trigonometric direction. 

The imaginary part of the resolvant 
\beq
\rho(z) = {1\over \beta\pi} M(z) \sqrt{(z-a)(b-z)}
\eeq
is the average density of eigenvalues of the random matrix (as an example, 
consider the gaussian case: $V$ is quadratic, i.e. $V'$ is of degree $d=1$, thus 
$M(z)$ is a constant and $\rho(z)=\sqrt{(z-a)(b-z)}$ is the famous semi-circle 
law).

\bigskip
\noindent {\bf Some notations:}

It will be convenient to parametrize $z$ as:
\beq\label{paramtrigo}
z= {a+b\over 2} + {b-a\over 2} \cos\phi \virg \alpha:={b-a\over 4}
\eeq
We write:
\beq\label{defsigma}
\sigma(z) := (z-a)(z-b) \virg \sqrt{\sigma(z)} = 2 i\alpha \sin\phi
\eeq

Note that $\phi(z)$ is a multi-valued function.
We will see that both determinations $\phi$ and $-\phi$ will enter the 
asymptotic expression of the orthogonal polynomials when $z\in [a,b]$.

\subsection{Variations of $W_0$ with respect to $T$ and $r$}

It can be proven (see \cite{Eynard:1997jf} for instance) that
\beq\label{WT}
W_T(z) := {\D \over \D T} T W_0(z) = {1\over \sqrt{\sigma(z)}} = {\D \phi(z) 
\over \D{z} }
\eeq
and
\beq\label{Wr}
W_r(z) := \beta {\D \over \D r} W_0(z) = -{1\over \sqrt{\sigma(z)}}{ 
\sqrt{\sigma(z)}-\sqrt{\sigma(x)} \over (z-x)} + {1\over \sqrt{\sigma(z)}}
\eeq
In particular at $z=x$, we have:
\beq\label{Wrxequalz}
W_r(x) =  -{ \sigma'(x) \over 2\sigma(x) } + {1\over \sqrt{\sigma(x)}}
\eeq

\subsection{Contribution of $W_1$}

For $T=1$ and $r=0$, and at order $O(1/n)$, the equation of motion reduces to:
\beq\label{eqmotionortho}
W^2(z) - {\eta\over 2n} W'(z) + O(1/n^2)  = {2\over \beta} V'(z) W(z) - Q(z) 
\eeq
and we expand $W(z)$ at first order in $1/n$ as:
\beq
W(z) \sim W_0(z) + {\eta\over 2n} W_1(z) + O(1/n^2)
\eeq
At order $1/n$, eq.\ref{eqmotionortho} gives (using the value of $W_0(z)$ from \ref{onecutsol}):
\beq
{2\over \beta} W_1(z) =  { Q_1(z) - W_0'(z)    \over M(z) \sqrt{\sigma(z)}}
\eeq
where $Q_1(z)$ is a polynomial of degree $d-2$.

Let us factorize $M(z)$ (recall that $d= {\rm deg} \, V'$ and $g$ is the leading coef. of $V'$):
\beq
M(z) = g \prod_{k=1}^{d-1} (z-z_k)
\eeq
and decompose $W_1$ in single pole terms.
The condition that $W_1(z)$ is regular when $z=z_k$ allows to determine the 
polynomial $Q_1(z)$, and eventually we get:

\beq\label{Wun}
 W_1(z) = {\sigma'(z)\over 4\sigma(z)} + {1\over 2}\sum_{k=1}^{d-1} { 
\sqrt{\sigma(z)}-\sqrt{\sigma(z_k)}\over (z-z_k)\sqrt{\sigma(z)}}
- {d\over 2\sqrt{\sigma(z)}}
\eeq
With the parametrization 
$z={a+b\over2} + 2\alpha \cos\phi $ and $z_k={a+b\over2} + 2\alpha 
\cos{\phi_k} $, we have
\beq\label{Wuntrigo}
W_1(z) = {\D\over \D{z}} \left[ {1\over 4}\ln{\sigma(z)} + \sum_{k=1}^{d-1} 
\ln{\sin{({\phi+\phi_k\over 2})}}
- { d\over 2}\phi  \right]
\eeq

\subsection{Asymptotics of the skew-orthogonal polynomials}

Eventually we have computed all the contributions to the asymptotics of the 
resolvant:
\beq
W(z) \sim {1\over T}W_0(z) + {T-1\over T} W_T(z) + {r\over \beta}  W_r(z) + 
{\eta\over 2n} W_1(z) +O(1/n^2)
\eeq
i.e.:
\beq
2n W(z) \sim \beta N_\beta W_0(z) + (2n-\beta N_\beta) W_T(z) + W_r(z) + \eta 
W_1(z) +O(1/n)
\eeq
where $W_0$, $W_T$, $W_r$, $W_1$ are given by \ref{onecutsol}, \ref{WT}, 
\ref{Wr} (or \ref{Wrxequalz}), \ref{Wun} (or \ref{Wuntrigo}).

\bigskip
\noindent Combining everything together, we get:

\bigskip

\noindent $\bullet$ $\beta=2$.
From $P'_n/P_n = n W(x)$, we get the asymptotic orthogonal polynomial (already 
known \cite{Brezin:1993,Eynard:1997jf,DeiftBook}):

\beq\label{asympPnU}
P^{(2)}_n(x)  \ee{-{N\over 2} V(x)} \sim  { C_n^{(2)} \over 
\sqrt{2i\alpha\sin\phi} } 
\,\, \ee{-Ni\pi \int_a^x \rho(y) \D{y} } \,\, \ee{i(n-N+{1\over 2}) \phi} 
\quad+\quad {\rm c.c.}
\eeq
The normalization constant $C_n^{(2)}=\alpha^{n+{1\over 2}}$ is such that 
$P_n(x)\sim x^n$ for large $x$.\\
\ref{asympPnU} is basically the contribution of $W_0$, which is the same for all three cases $\beta=1,2,4$.
The $\beta=1$ and $\beta=4$ cases contain an extra contributions from $W_1$.

\bigskip

\noindent $\bullet$ $\beta=1$.
From $P'_{2n}/P_{2n} = 2n W(x)$ we get:
\beq\label{PnevenOasymp}
\encadremath{
P^{(1)}_{2n}(x)  \ee{-{N} V(x)} \sim  { C_n^{(1)}  \over \sqrt{2i\alpha\sin\phi} 
} \,\, 
\ee{-Ni\pi \int_a^x \rho(y) \D{y} } \,\, \ee{i(2n+1-N-{d\over 2}) \phi} \,\, 
M_+(\phi)  \quad+\quad {\rm c.c.}
}
\eeq
where
\beq
M_+(\phi) =M_-(-\phi) = \prod_{k=1}^{d-1} 2i\, \sin{\left({\phi+\phi_k\over 
2}\right)}
\eeq
note that $M(x) = g \alpha^{d-1} M_+(\phi) M_-(\phi) $,
where $g$ is the leading coefficient of $V'(x)$.
\beq
C_n^{(1)} = \alpha^{2n+{1\over 2}} \prod_{k=1}^{d-1} \ee{-i\phi_k/2}
\eeq
is the normalization constant chosen so that $P_{2n}(x) \sim x^{2n}$ for large 
$x$.

The odd polynomial is found from $P_{2n+1}/P_{2n} = <x+\Tr M + c_n>$ and \hbox{$<\Tr\ M> = 2n\mathop{\rm lim}_{z\to\infty} z^2(W(z)-1)$} (note that we need \ref{Wr}, not \ref{Wrxequalz}).
The whole $x$ dependance of $P_{2n+1}/P_{2n}$ comes from $x+  \mathop{\rm 
lim}_{z\to\infty} z^2(W_r(z)-1) = \sqrt{\sigma(x)}-{a+b\over 2}$.
Therefore, (and up to an arbitrary linear combination with $P_{2n}$), we have:

\beq\label{PnoddOasymp}
\encadremath{
P^{(1)}_{2n+1}(x)  \ee{-{N} V(x)} \sim  C_n^{(1)} \,\, \sqrt{2i\alpha\sin\phi}  
\,\, 
\ee{-Ni\pi \int_a^x \rho(y) \D{y} } \,\, \ee{i(2n+1-N-{d\over 2}) \phi} \,\, 
M_+(\phi)  \,+\, {\rm c.c.}
}
\eeq

\bigskip

\noindent $\bullet$ $\beta=4$.
From $P'_{2n}/P_{2n} = 2n W(x)$ we get:
\beq\label{PnevenSpasymp}
\encadremath{
P^{(4)}_{2n}(x)  \ee{-{N\over 2} V(x)} \sim  {  C_n^{(4)}  \over 
\sqrt{2i\alpha\sin\phi} } \,\, \ee{-2Ni\pi \int_a^x \rho(y) \D{y} } \,\, 
\ee{i(2n+1-2N+{d\over 2}) \phi} 
\,\, {M_-(\phi)\over i\rho(x)}  \,+\, {\rm c.c.}
}
\eeq
with normalization constant:
\beq
C_n^{(4)} = {g\over 4\pi} \alpha^{2n+d+{1\over 2}} \prod_{k=1}^{d-1} 
\ee{i\phi_k/2}
\eeq
and
\beq\label{PnoddSpasymp}
\encadremath{
 P^{(4)}_{2n+1}(x)  \ee{-{N\over 2} V(x)} \sim  C_n^{(4)}  \, 
\sqrt{2i\alpha\sin\phi} 
\,\, \ee{-2Ni\pi \int_a^x \rho(y) \D{y} } \,\, \ee{i(2n+1-2N+{d\over 2}) \phi} 
\,\, {M_-(\phi)\over i\rho(x)}  \,+\, {\rm c.c.}
}
\eeq
Note that we have used:
$i\rho(x)={g\over 4\pi} \alpha^{d}\, M_+(\phi)\, M_-(\phi) 
\,\, 2i\sin\phi $.

\bigskip

{\bf \noindent Some remarks:}

\noindent - the derivation presented here is actually valid only when $x\notin 
[a,b]$, giving only one exponential term, with the determination of $\phi(x)$ 
(from \ref{paramtrigo}) such that $P_n(x)\ee{-NV(x)}$ decreases when 
$x\to\infty$.
When $x\in [a,b]$, a carefull analysis shows that both determinations of 
$\phi(x)$ must be taken into account.
The only effect is to add the complex conjugate exponential (c.c.) to the 
asymptotics, so that $P_n(x)$ is indeed real when $x\in [a,b]$.
Outside $[a,b]$, $P_n\ee{-NV}$ decreases exponentially, and in $[a,b]$, it 
oscillates like a cosine function, and it indeed has $n$ zeroes.

\noindent - Our derivation was carried out only in the "one-cut" case. We have 
assumed that the support of $\rho(x)$ is connected and is made of one interval 
$[a,b]$.

\noindent - Those asymptotics are not valid when $x$ is close to $a$ or $b$.

\noindent - Note that the above expressions all have the correct large $x$ 
behaviour: $P_{n}(x) \sim x^{n} $.
It can be seen easily if one remembers that $x\sim \alpha\ee{i\phi}$ when 
$i\phi\to 
+\infty$.

\subsection{Check of orthogonality}

We have presented a derivation of the asymptotics 
(\ref{PnevenOasymp}-\ref{PnoddSpasymp}), so that there should be no reason to 
doubt they fulfill the orthogonality condition.
However, it is interesting to see how.
We will just sketch the procedure:\\
In all cases, we have to compute integrals of $P_n P_m \ee{-NV}$, with $x$ 
running 
from $-\infty$ to $+\infty$.
The contributions outside $[a,b]$ are exponentially small, the integrals 
can thus be computed inside $[a,b]$.
Within $[a,b]$, terms which oscillate exponentially fast like $\ee{Ni\pi 
\int\rho}$, average to zero at order $O(1/N)$, so that at leading order, it is 
sufficient to 
consider only the cross-terms in the product $P_n P_m$, with opposite signs for 
the two determinations of $\phi$.\\
In the $\beta=1$ case, the scalar product $<P_n | P_m>$ of \ref{scalarO} can be 
computed by integration by part.
For that, you need a primitive of $P_n \ee{-NV}$, which is achieved at leading 
order by dividing \ref{PnevenOasymp} or \ref{PnoddOasymp} by $\rho(x) = {\rm 
cte}\,\, M_+(\phi) M_-(\phi) \sin\phi $.\\
In the $\beta=4$ case, you need a derivative of $P_n \ee{-{N\over 2}V}$, which is 
achieved at leading order by multiplying \ref{PnevenSpasymp} or 
\ref{PnoddSpasymp} by $\rho(x) = {\rm cte}\,\, M_+(\phi) M_-(\phi) \sin\phi $.

Then you find that in both cases ($\beta=1$ and 4), and up to unimportant 
constant factors, you have at leading order in $1/n$:

\beq
<P_{2n} | P_{2m}>  \propto \int_0^{\pi} \D\phi \,\, {\sin{2(n-m)\phi}\over 
\sin\phi} =  0
\eeq

\beq
<P_{2n+1} |  P_{2m+1}>  \propto  \int_0^{\pi} \D\phi \,\,\sin\phi 
\,\sin{2(n-m)\phi}  =   0
\eeq

\beq
<P_{2n+1} |  P_{2m}>   \propto   \int_0^{\pi} \D\phi \,\, \cos{2(n-m)\phi} 
\propto  
 \,\,\delta_{nm}
\eeq
which confirms that our asymptotics indeed fulfill the orthogonality properties.

Taking into account properly the constant factors, we can determine the 
$h_n$'s:

\begin{itemize}
\item $\beta=2$: 
\beq
h_n^{(2)} \sim 2\pi \,\,\alpha^{2n+1}
\eeq

\item $\beta=1$: 
\beq
h_n^{(1)} \sim {16\pi\over Ng\alpha^{d+1}} \,\,\alpha^{4n+3}
\eeq

\item $\beta=4$: 
\beq
h_n^{(4)} \sim 2N \pi g\alpha^{d+1} \,\,  \alpha^{4n+1}
\eeq

\end{itemize}

\section{Conclusions}

Therefore, we have obtained some exact integral expressions and asymptotics for 
the skew-orthogonal polynomials involved in the Orthogonal and Symplectic random 
matrix ensembles.

Our asymptotics were derived in the "one-cut" case only, though it seems obvious 
that the result could be extended easily to the multicut case, following the 
method of \cite{BoEyDa00} or \cite{Deift:1999}, it would involve 
hyper-elliptical 
theta-functions instead of exponentials.

Another possible extension of the method presented here is to "multi-matrix 
models", and a time dependant matrix, as in \cite{Eynard:1997jf}.
It seems that the same kind of asymptotics could be obtained.

\bigskip

The asymptotics of the skew-orthogonal polynomials are useful to evaluate the 
kernels:
\beq\label{kerneldef}
K(\l,\mu) = {1\over N} \sum_{n=0}^{N-1} {1\over h_n} \left( P_{2n}(\l) 
P_{2n+1}(\mu) - P_{2n+1}(\l) P_{2n}(\mu) \right) \,\ee{-NV(\l)}\, \ee{-NV(\mu)}
\eeq
which give all the correlation functions.
For instance with $\beta=4$, we have \cite{MehtaBook}
\bea
\rho(\l) & = & -\left. {\d\over \d\l}K(\l,\mu)\right|_{\mu=\l} \\
\rho_c(\l,\mu) & = & - {\d\over \d\l}K(\l,\mu) \,\, {\d\over \d\mu}K(\l,\mu) + 
K(\l,\mu) \,\, {\d\over \d\l}{\d\over \d\mu}K(\l,\mu)
\eea
In order to use the asymptotics of the orthogonal polynomials in 
\ref{kerneldef}, one needs a generalization of the Darboux-Christoffel theorem, 
which allows to write $K(\l,\mu)$ in terms of a few $P_n$ only with $n$ close to the Fermi level $n_F$.
With asymptotics of the type \ref{asympPnU}, \ref{PnevenOasymp}, 
\ref{PnoddOasymp}, \ref{PnevenSpasymp} or \ref{PnoddSpasymp}, the 
Darboux-Christoffel Theorem merely amounts to a formal resummation of the 
geometrical series (it was 
proven in \cite{Eynard:1997jf} for hermitian multi-matrix models, and we cannot see any 
reason why the same proof would not work here). For instance in the 
$\beta=4$ case, the generalization of the Darboux-Christoffel theorem reads:
\beq
\sum_{n=0}^{N-1} \ee{i(2n+3-2N)(\phi(\l)-\phi(\mu))} \sim {1\over 2i 
\sin{(\phi(\l)-\phi(\mu))} }
\eeq
This trick allows to find asymptotics for the kernels $K(\l,\mu)$, and then 
asymptotics for all the correlation functions.
One can then easilly check that  in the short distance regime $|\l-\mu|\sim 
O(1/N)$, the universal 2-point connected correlation function is well reproduced 
, and that in the long distance regime $|\l-\mu|\sim O(1)$, the smoothed 2-point 
connected correlation function is correctly reproduced too.
The leading behaviour of short and long distance correlation functions was 
already known from other methods \cite{MehtaBook}, so that our method does not provide any new 
result for the correlation functions.
However, it seems that our asymptotics can be used to build a rigorous 
mathematical proof of the universality, following the method of 
\cite{Pastur:1997}, 
because they allow a good control of the approximations.

\smallskip

In addition, the fact that the skew-orthogonal polynomials are exactly the 
average characteristic polynomials of the random matrices is remarkable.
It would be interesting to understand the generality of this result, and for 
instance try to generalize it to the other random matrix ensembles related to 
Cartan's classification of symmetric spaces \cite{Zirnbauer:1996,Caselle:1996fy}.

\hfill\eject
\bibliographystyle{prsty}
\bibliography{polbib,biblio/RMTcondmat,biblio/RMTGQ,biblio/eynard,biblio/condmat,biblio/strings,biblio/QCDRMT,biblio/RMTespacessym}

\end{document}